\documentclass{appolb}
\usepackage{graphicx}
\usepackage{amsmath,amssymb}    % need for subequations
\usepackage{graphicx}   % need for figures
\usepackage{color}
\usepackage{cancel}
\usepackage{siunitx}  % for unit formating
\usepackage[colorlinks,linkcolor=blue,citecolor=blue]{hyperref}   % use for hypertext links, including those to external documents and URLs
\usepackage{bm} % for bold caligraphic font
\usepackage{import}
\usepackage{ulem}

\def\L{{\mathcal{L}}}
\def\P{{\mathcal{P}}}

\def\x{{\bm x}}

\def\p{{\bf p}}
\def\rmd{{\rm d}}

\newcommand\nda{\end{align}}
\def\Eqs#1{Eqs.~(\ref{#1})}

\advance\parskip 1.9pt
\advance\voffset -0.2in

\def\be{\begin{equation}}
\def\ee{\end{equation}}
\def\bg{\begin{eqnarray}}
\def\nd{\end{eqnarray}}

\newcommand{\beq}{\begin{eqnarray}}
\newcommand{\eeq}{\end{eqnarray}}

\begin{document}
% \eqsec  % uncomment this line to get equations numbered by (sec.num)
\title{Emergence of hydrodynamics in expanding relativistic plasmas%
\thanks{Presented at \it{Excited QCD 2022}, Sicily, Italy, 23-29 October, 2022.}%
% you can use '\\' to break lines
}
\author{Jean-Paul Blaizot, \address{Institut de Physique Th{\'e}orique, Universit\'e Paris Saclay, 
        CEA, CNRS, 
	F-91191 Gif-sur-Yvette, France} 
}
\maketitle
\begin{abstract}
I consider a simple set of equations that govern the expansion of boost-invariant plasmas of massless particles.  These equations describe the transition from a collisionless regime at early time to hydrodynamics at late time. Their mathematical structure encompasses all versions of second order hydrodynamics.  We  emphasize that the apparent success of  Israel-Stewart hydrodynamics at early time has little to do with ``hydrodynamics'' proper,  but rather with a particular feature of Israel-Stewart equations that allows them to effectively mimic the collisionless regime. 
\end{abstract}

\section*{}
In this note, I consider an idealization of  the early stages of a high-energy heavy-ion collision, where the produced matter  expands longitudinally along the collision axis in a boost invariant fashion, undergoing the so-called Bjorken expansion \cite{Bjorken:1982qr}.  The matter is supposed to occupy uniformly the plane transverse to the collision axis (the $z$-axis). The discussion will be based on the simple kinetic equation \cite{Baym:1984np}, 
  \be
  \left[ \partial_\tau-\frac{p_z}{\tau}\partial_{p_z} \right]f(\p,\tau)=-\frac{f(\p,\tau)-f_{\rm eq}(p/T)}{\tau_R},
  \ee  
  where $f$ denotes a distribution function for massless particles, and the right-hand side is a collision term treated in the relaxation time approximation ($f_{\rm eq}(p/T)$ is the local equilibrium distribution function). 

In the case of massless particles, the energy momentum tensor has two independent components, which can be identified to the energy density $\varepsilon$ and the difference between the longitudinal and transverse pressures $\P_L-\P_T$. These two quantities are special moments of the distribution function,  $\varepsilon=\L_0$ and $\P_L-\P_T=\L_1$, where for any integer $n$, we  define \cite{Blaizot:2017lht} 
  \be 
\L_n\equiv \int \frac{d^3 \p}{(2\pi)^3} \,p^2 P_{2n}(p_z/p) f_p(t,\x, \p), \ee
 with $P_n(x)$ a Legendre polynomial and $p=|\p|$  \footnote{These moments $\L_n$, introduced in \cite{Blaizot:2017lht}, are distinct from those most commonly used  (see e.g. \cite{Strickland:2018ayk}). They also differ slightly from those used in \cite{Behtash:2019txb}. Note that although the knowledge of the $\L_n$ moments does not allow us to reconstruct from them the distribution function, they provide an exact description of the components of the energy-momentum tensor. }.
   Owing to the symmetries of the Bjorken expansion, the moments $\L_n $ depend only on the proper time $\tau=\sqrt{t^2-z^2}$. They obey the coupled equations  \cite{Blaizot:2017ucy}
\begin{subequations}
\label{eq:Lequ}
\begin{align}
\label{eq:Lequa}
\frac{\partial \L_0}{\partial \tau} =& - \frac{1}{\tau}(a_0 \L_0 + c_0 \L_1)\,, \\
\label{eq:Lequb}
\frac{\partial \L_1}{\partial \tau} =& - \frac{1}{\tau}(a_1 \L_1 + b_1 \L_0+{c_1\L_2}) - \frac{\L_1}{\tau_R}.
\end{align}
\end{subequations}
The  coefficients, $a_0=4/3$, $a_1=38/21$, etc,  are pure numbers whose values are fixed by the geometry of the expansion. The last term in Eq.~(\ref{eq:Lequb}), proportional to the collision rate $1/\tau_R$,  isolates in a transparent way the effect of the collisions. Without this term, Eqs.~(\ref{eq:Lequ})  describe free streaming. In this regime, the moments evolve as power laws governed by the eigenvalues of the linear  system. The collision term in Eq.~(\ref{eq:Lequb}) produces a damping of  $\L_1$ and drives the system towards isotropy, a prerequisite for local equilibrium. When $\L_1=0$, the system behaves as in ideal hydrodynamics $\L_0\sim \tau^{-a_0}$. There is no contribution of the collision term in  Eq.~(\ref{eq:Lequa}) since collisions conserve energy. 
 The $\L_n$ moments have all the same dimension, that of the energy density. Eqs.~(\ref{eq:Lequ}) are the first in an infinite hierarchy of equations that couple $\L_n$ to its nearest neighbours, $\L_{n+1}$ and $\L_{n-1}$.  Thus, in Eqs.~(\ref{eq:Lequ}) $\L_1$ is coupled to $\L_0$ and $\L_2$.  After an appropriate treatment of $\L_2$,  Eqs.~(\ref{eq:Lequ})  yield an effective theory for  $\L_0$ and $\L_1$, that is for the energy momentum tensor. In particular these equations contain ``second order'' hydrodynamics as a special limit.

To see that, we express the moments in terms of the more familiar hydrodynamical variables. We call ${\cal P}$ the equilibrium pressure (related to the energy density by the equation of state), and  set $\pi=-c_0 \L_1$ with $\pi$ the viscous pressure. Then, Eq.~(\ref{eq:Lequa}) takes the form
\be\label{eq:energy_expansion}
\frac{\rmd \varepsilon}{\rmd \tau}+\frac{\varepsilon+\P}{\tau} =\frac{\pi}{\tau}.
\ee
This equation  translates the conservation of the energy momentum tensor, $\partial_\mu T^{\mu\nu}=0$, for Bjorken flow. In ideal hydrodynamics, the viscous pressure is neglected ($\L_1\to 0$), and, for massless particles,  ${\cal P}=\varepsilon/3$. The solution of Eq.~(\ref{eq:energy_expansion}) is then $\varepsilon(\tau)\sim \tau^{-4/3}$. 
By taking into account the viscous effects via the leading order constitutive equation $\pi=4\eta/(3\tau)$, with $\eta$ the shear viscosity, one obtains the  Navier-Stokes equation:
\be
\frac{\rmd \varepsilon}{\rmd \tau} =- \frac{a_0}{\tau}\left( \varepsilon -\frac{\eta}{\tau}\right).
\ee

An equation similar to 
 Eq.~(\ref{eq:Lequb}) was introduced by Israel and Stewart \cite{Israel:1979wp} in order to cure problems of the relativistic Navier-Stokes equation. In the present context, it takes the form of a relaxation equation for the viscous pressure  $\pi$, forcing it to relax towards its Navier-Stokes value  $4\eta/(3\tau)$ over a time scale $\tau_\pi$:
 \be
\partial_\tau \pi+\frac{a_1^{IS}}{\tau}\pi=-\frac{1}{\tau_\pi} \left( \pi-\frac{4\eta}{3\tau}  \right).
\ee
This equation reduces identically to Eq.~(\ref{eq:Lequb}) after setting  $\tau_\pi=\tau_R$ and $ a_1^{IS}=a_1$.  It can be verified that all second order formulations of hydrodynamics for the boost invariant system share the same mathematical structure as that encoded in the linear system (\ref{eq:Lequ}), modulo the adjustment of the parameters $b_1$ (or $\eta$), $\tau_R\to \tau_\pi$ and $a_1-a_0\to \lambda_1$, where $\tau_\pi$ an $\lambda_1$ may be viewed as second order transport coefficients (see  \cite{Blaizot:2021cdv} for a more complete discussion).

To proceed further, it is convenient to define 
\be
g(w) \equiv \frac{\tau}{\L_0}\frac{\partial \L_0}{\partial \tau}=-1-\frac{\P_L}{\varepsilon},
\ee
where $w\equiv \tau/\tau_R$. 
The quantity $g(w)$ may be viewed as the exponent of  the power laws obeyed by the energy density at early or late times (in both cases $g(w)$ becomes constant). It is also a measure of the pressure asymmetry. In particular, the second relation, which follows easily from Eqs.~(\ref{eq:Lequ}), shows that in the free streaming regime where $\P_L=0$, $g=-1$, while in the hydrodynamical regime where $\P_L=\varepsilon/3$, $g=-4/3$.  In terms of $g(w)$, \Eqs{eq:Lequ} become a first order nonlinear ODE\footnote{An equation very similar to this one was considered in \cite{Heller:2015dha}}
\beq
\label{eq:dif_gw}
&& w\frac{\rmd g}{\rmd \ln w}=\beta(g,w),\nonumber\\
&&-\beta(g,w)=g^2+\left(a_0+a_1+w\right)g+a_1a_0-c_0b_1+ a_0 w-c_0 c_1 \frac{\L_2}{\L_0}.\nonumber\\
\eeq
Let us first ignore the term $\L_2$. Then, in the absence of collisions, or for small $w$, this non linear equation has two fixed points, that we refer to as unstable ($g_-$) and stable ($g_+$) free streaming fixed points, whose values coincide with the eigenvalues of the linear system (\ref{eq:Lequ}) (with  $\L_2=0$). Numerically, $g_+=-0.929$, $g_-=-2.213$. As discussed in  \cite{Blaizot:2019scw} this fixed point structure is little affected when higher moments are taken into account, leading eventually to the exact values of the fixed points, respectively -1 and -2. In fact, to obtain an accurate description of the solution in the vicinity of a fixed point, it is enough to inject in Eq.~(\ref{eq:dif_gw}) the value of $\L_2$ in the vicinity of the corresponding fixed point, and this is known. For instance, near the stable free streaming fixed point  $\L_n/\L_0=A_n$, where $A_n$ is a known number (e.g. $\L_2/\L_0=3/8$). The effect of the entire tower of higher moments can then be absorbed in a renormalisation of the parameter $a_1$ of the two moment truncation:
\beq\label{eq:a1prime}
a_1\mapsto a_1'=a_1+c_1\frac{A_2}{A_1}=\frac{31}{15}.
\eeq 
With this value of $a_1$, the stable free streaming fixed point is exactly reproduced, i.e., $g_+=-1$.

This fixed point structure continues to play a role when collisions are switched on \cite{Blaizot:2019scw}: The unstable fixed point moves to large negative values, while the stable fixed point $g_+$ evolves adiabatically to the hydrodynamic fixed point, $g_*= -4/3$. The location of this ``pseudo fixed point'' as $w$ runs from $0$ to $\infty$ corresponds (approximately) to what has been  dubbed ``attractor''  \cite{Heller:2015dha}.  More precisely, the attractor is to be understood  as the particular solution of Eq.~(\ref{eq:dif_gw}), $g_{\rm att}(w)$,  that connects  $g_+$ as $w\to 0$ to $g_*$ as $w\to\infty$.  Such an attractor is made of three parts: the vicinities of the two fixed points, and the transition region. The two fixed points  are associated with different, well identified,  physics: one corresponds to hydrodynamics, the other to a collisionless regime. The vicinities of these fixed points can be described by viscous hydrodynamics for the first one, and perturbation theory for the second. The transition region requires information on both fixed points to be accurately accounted for.

From this perspective, the often used terminology of ``hydrodynamic attractor'' appears misleading. The gradient expansion is divergent, and the full solution of the kinetic equation can be obtained in terms of trans-series \cite{Heller:2015dha}. In such trans-series, the first non trivial correction to the hydrodynamic gradient expansion requires information about the early time dynamics (for an analytic solution of the system with $\L_2=0$ see \cite{Blaizot:2020gql})). This  information is necessary to control accurately the transition region between the two fixed points, that is to get a good description of the attractor.

%uncomment the following lines to place a figure
\begin{figure}[htb]
\centerline{%
\includegraphics[width=7cm]{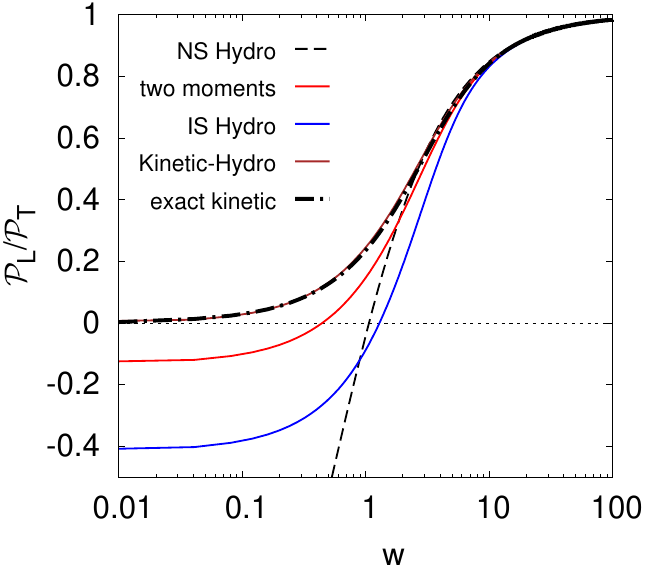}}
\caption{Plot of the attractor solution for the pressure ratio $\P_L/\P_T$ as a function of $w=\tau/\tau_R$. The dashed curve represents the solution of the Navier-Stokes equation. The curves labelled ``IS Hydro'', ``two moments'', ``Kinetic-Hydro'' correspond to different values of $a_1$, respectively 4/3, 31/28, 31/15. From \cite{Blaizot:2021cdv}. }
\label{Fig:fig1}
\end{figure}
%\emph{Acknowledgements.} --- 
%L.Y. is supported in part by National Natural Science Foundation of China (NSFC) 
%under Grant No. 11975079.

We have emphasized earlier the role of the higher moments in the determination of the free streaming fixed points, and indicated that in the vicinity of the stable fixed point this boils down to a renomalisation of the parameter $a_1$. 
Within Israel-Stewart theory, changing $a_1$ looks like changing a second order transport coefficient. However, in the vicinity of the hydrodynamic fixed point  the gradient expansion yields $\L_{n>1}\simeq 1/\tau^n$, so that $\L_2$ does not affect the hydrodynamic fixed point, nor its leading order viscous correction. The correct interpretation of changing $a_1$ is to put  the stable free streaming fixed point at its right place, and this has a strong impact on the whole attractor, except in the vicinity of the hydrodynamic fixed point.  This is clearly illustrated in Fig.~\ref{Fig:fig1}.  

It follows from this analysis that hydrodynamic behavior emerges where it is supposed to do so, namely when the collision rate becomes comparable to the expansion rate (i.e. when $\tau\gtrsim \tau_R$). The fact that Israel-Stewart equations apparently allow ``hydrodynamics''  to work at early time has little to do with hydrodynamics proper, but rather with the fact that the structure of Israel-Stewart  equations is similar to that of the moments of the kinetic equations. Thus, they capture features of the collisionless regime (but only approximately, unless $a_1$ is carefully adjusted - see in Fig.~\ref{Fig:fig1} the negative longitudinal  pressure obtained when $a_1$ differs from its proper value).

\bibliographystyle{unsrt}
\bibliography{refsbib}
\end{document}